\documentclass[12pt,a4paper]{article}

\usepackage{epsfig}

\begin{document}
\vspace*{-3cm}
\begin{flushright}
hep-ph/0307176 \\
July 2003
\end{flushright}
\vspace{0.5cm}
\begin{center}
\begin{Large}
{\bf Indirect signals of quark singlets at colliders}
\end{Large}

\vspace{0.5cm}
J. A. Aguilar--Saavedra \\
{\it Departamento de F\'{\i}sica and CFIF, \\
Instituto Superior T\'ecnico, P-1049-001 Lisboa, Portugal}
\end{center}

\begin{abstract}
In this talk we review some of the most relevant effects that the mixing with
a $Q=2/3$ quark singlet might give at high energy colliders and $K$ and $B$
factories.
\end{abstract}

With the new colliders under construction or planned, the Standard Model (SM)
sill be tested with an unprecedented precision in the forthcoming years. The
precise measurement of the properties of the SM particles may shown indirect
evidence of new physics, complementing the direct signals given by the
observation of new particles. Our aim here is to describe some of the indirect
effects that could be observed as a consequence of the mixing with quark
singlets. These new ``exotic'' quarks have both their left- and right-handed
components transforming as singlets under $\mathrm{SU}(2)_L$. Apart from their 
theoretical motivations \cite{papiro1,papiro1b}, there is a clear
phenomenological reason
for the study of models with quark singlets. These models are the simplest ones
in which the unitarity of the $3 \times 3$ Cabibbo-Kobayashi-Maskawa (CKM)
matrix is broken, and large deviations in top quark couplings with respect to
the SM predictions are possible. In other SM extensions, like two Higgs doublet
models or supersymmetry, the $3 \times 3$ CKM matrix is unitary and hence the
top charged couplings have the same values predicted by the SM.

For illustration we consider a SM extension with an extra $Q=2/3$ quark singlet.
The most important effects of the mixing are caused by the modification of the
SM weak interaction Lagrangian due to the mixing with the singlet. In this
model, the weak interaction terms read
\begin{eqnarray}
{\mathcal L}_W & = & - \frac{g}{\sqrt 2}
\,\bar u_L \gamma^\mu V d_L \,W_\mu^+ +\mathrm{h.c.} \,, \nonumber \\
{\mathcal L}_Z & = & - \frac{g}{2 c_W} \left(
\bar u_L \gamma^\mu X^u u_L - \bar d_L \gamma^\mu d_L
  - 2 s_W^2 J_\mathrm{EM}^\mu \right) Z_\mu \,,
\end{eqnarray}
with $V$ the CKM matrix, in this case of dimension $4 \times 3$, and $X^u = V
V^\dagger$ a $4 \times 4$ hermitian matrix, not necessarily diagonal. This
modification of the charged and neutral current terms leads to new contributions
to precision observables, like the ratios $R_b$, $R_c$, the forward-backward
asymmetries measured at LEP and SLD and the oblique parameters. Moreover, there
are important effects on meson physics, for instance in $K$, $D$ and $B$
oscillations, in the decays $b \to s \gamma$, $b \to s \, l^+ l^-$ and in the
rare
$K$ decays $K^+ \to \pi^+ \nu \bar \nu$, $K_L \to \mu^+ \mu^-$. The agreement of
this model with experiment implies that the charged current couplings of the new
mass eigenstate $V_{Td}$, $V_{Ts}$ and the nondiagonal terms in $X^u$ must be
relatively small, of order $O(10^{-1})$ at most. However, it is possible to have
a sizeable mixing with the third generation, with observable effects at present
and future colliders \cite{papiro2}.

\begin{figure}[htb]
\begin{center}
\mbox{\epsfig{file=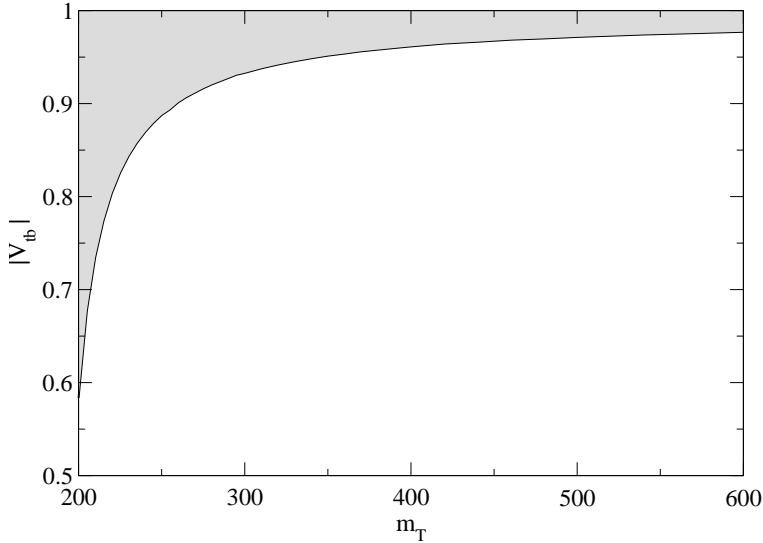,width=10cm,clip=}}
\end{center}
\caption{Allowed values of $|V_{tb}|$ (shaded area) as a function of
the mass of the new quark.
\label{fig:1} }
\end{figure}

One remarkable effect of the mixing is the decrease of $V_{tb}$ from the SM
prediction $V_{tb} = 0.999$. The lower limit on $V_{tb}$ depends on the mass of
the new quark $m_T$, and can be as small as $V_{tb} \geq 0.6$, as can be
observed in Fig.~\ref{fig:1}. Such deviations could be easily noticed in the
measurement of $V_{tb}$ in single top production at LHC \cite{papiro3}. On the
other hand, the $Z$ diagonal coupling $X_{tt}^u = |V_{td}|^2 + |V_{ts}|^2
+ |V_{tb}|^2 \simeq |V_{tb}|^2$ is closely related to $V_{tb}$, and its
measurement in top pair production at TESLA might be more precise.

\begin{figure}[htb]
\begin{center}
\mbox{\epsfig{file=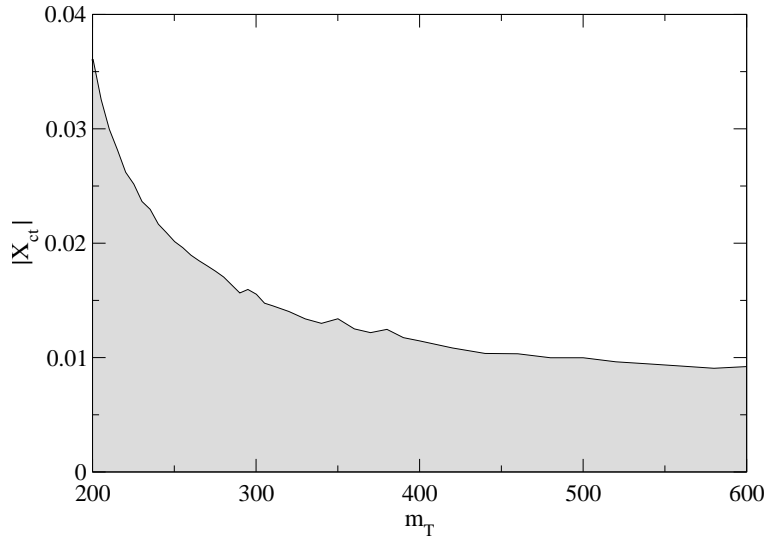,width=10cm,clip=}}
\end{center}
\caption{Allowed values of the FCN coupling $|X_{ct}|$ (shaded area)
as a function of the mass of the new quark.
\label{fig:2} }
\end{figure}

Other striking effect is the appearance of tree-level flavour-changing neutral
couplings of order $O(10^{-2})$ between $Q=2/3$ quarks \cite{papiro7},
which could lead to flavour-changing top decays and single top production at LHC
\cite{papiro4} and TESLA \cite{papiro5} at observable rates
(see Fig.~\ref{fig:2}).


The mixing with a singlet may also yield new effects in low energy observables.
For instance, the branching ratio of the theoretically clean decay $K_L \to
\pi^0 \nu \bar \nu$ can reach $4.4 \times 10^{-10}$, more than one order of
magnitude above the SM prediction and readily observable in planned experiments
\cite{papiro6}. Other interesting quantity is the mass difference in the $B_s$
system, which can range between 13.1 and 37.7 ps$^{-1}$ (the SM expectation is
17.6 ps$^{-1}$). Another excellent place to look for new physics is through CP
asymmetries in $B$ decays. A good example is the time dependent CP asymmetry in
the decay $B_s \to D_s^+ D_s^-$. This asymmetry is almost free of penguin
pollution and it is predicted to be close to zero in the SM, but can have any
value between $-0.3$ and $0.4$ in this model.

In summary, the existence of a new quark singlet can have interesting
consequences at large colliders and at $K$ and $B$ factories. Despite the strong
constraints given by present experimental data, there is still large room for
new effects, which might show up in addition to the direct observation of a new
quark.

\end{document}